\documentclass[12pt]{article}

\begin{document}

\title{\textbf{Parallel and Orthogonal Cylindrically Symmetric Self-Similar Solutions}}

\author{M. Sharif \thanks{msharif@math.pu.edu.pk} and Javaria Zanub \\
\\Department of Mathematics, University of the Punjab,\\Quaid-e-Azam
Campus Lahore-54590, Pakistan.}
\date{}

\maketitle
\begin{abstract}
In this paper, we evaluate kinematic self-similar perfect fluid and
dust solutions for the most general cylindrically symmetric
spacetime. We explore kinematic self-similar solutions of the first,
second, zeroth and infinite kinds for parallel and orthogonal cases.
It is found that the parallel case gives solutions for both perfect
fluid and dust cases in all kinds except the zeroth kind of the dust
case where there exists no solution. The orthogonal perfect fluid
case gives stiff fluid solution only in the first kind and vacuum
solution for the dust case. We obtain a total of thirteen solutions
out of which eleven are independent. The correspondence of these
solutions with those already available in the literature is also
given.
\end{abstract}


\section{Introduction}

Self-similarity simplifies the mathematical complexity of partial
differential equations (PDEs). For an appropriate matter field, a
set of field equations remains unchanged under a scale
transformation. The solutions which are invariant under scale
transformation are known as self-similar solutions. The special
feature of these solutions is that, by a suitable coordinate
transformations, the number of independent variables can be reduced
by one and hence reduces the PDEs into ordinary differential
equations (ODEs). This variable, a dimensionless combination of the
independent variables, is called self-similar variable.

Einstein's field equations (EFEs) are highly non-linear, second
order PDEs. In order to solve these equations, several kinds of
symmetry restrictions have been imposed by different people.
Self-similarity is one of these restrictions used to obtain the
exact solutions of the EFEs. In Newtonian gravity, self-similar
solutions have been investigated by many authors to obtain the
realistic solutions of gravitational collapse leading to star
formation [1]. However, in General Relativity, self-similar
solutions were first studied by Cahill and Taub [2] which correspond
to Newtonian self-similarity of the homothetic class (also called
first kind). Carter and Henriksen [3,4] defined the self-similarity
of second, zeroth and infinite kinds.

McIntosh [5] found that a stiff fluid $k = 1$ is the only compatible
perfect fluid with the homothety in the orthogonal case. Benoit and
Coley [6] studied analytic spherically symmetric solutions of the
EFEs coupled with a perfect fluid and admitting a kinematic
self-similar (KSS) vector of the first, second and zeroth kinds.
Carr et al. [7] investigated solution space of self-similar
spherically symmetric perfect fluid models and physical aspects of
these solutions. They combined the state space description of the
homothetic approach with the use of the physically interesting
quantities arising in the co-moving approach. Coley and Goliath [8]
discussed self-similar spherically symmetric cosmological models
with a perfect fluid and a scalar field with an exponential
potential. Sintes et al. [9] considered spacetime which admits a KSS
vector of the infinite kind without using the equation of state
(EOS).

Maeda et al. [10-12] investigated the KSS vector of the second,
zeroth and infinite kinds in the tilted, orthogonal and parallel
perfect fluid cases for spherically symmetric spacetime. Chan et al.
[13] studied the static vacuum plane symmetric spacetimes with two
non-trivial solutions: the Taub solution and the Rindler solution.
The same authors [14] studied the spherical gravitational collapse
of a compact packet consisting of perfect fluid. They concluded that
when the collapse has continuous self-similarity, the formation of
black holes always starts with zero mass and when collapse has no
self-similarity, the formation of black holes always starts with a
finite non-zero mass.

In recent papers [15-17], Sharif and Aziz have explored the KSS
solutions of a special cylindrically symmetric, special plane
symmetric and the most general plane symmetric spacetimes. This
analysis has been extensively given for the perfect fluid and dust
cases with tilted, parallel and orthogonal vectors. The same authors
[18-20] have also studied the physical properties of such solutions
for spherically, cylindrically and plane symmetric spacetimes
respectively. In this paper, we investigate the KSS solutions of the
most general cylindrically symmetric spacetimes for both perfect
fluid and dust cases. We explore the KSS solutions of the first,
second, zeroth and infinite kinds for the parallel and orthogonal
cases.

The paper has been organized as follows. In section \textbf{2}, we
discuss kinematic self-similarity for cylindrically symmetric
spacetimes. Sections \textbf{3} and \textbf{4} are devoted to the
parallel perfect fluid and dust solutions respectively. The
orthogonal perfect fluid and dust solutions are investigated in
section \textbf{5}. Finally, we summarize and discuss the results in
section \textbf{6}.

\section{Cylindrically Symmetric Spacetimes and Kinematic Self-Similarity}

The most general cylindrically symmetric spacetime is given in the
form [21]
\begin{equation}
ds^{2}=e^{2\nu(t,r)}dt^{2}-e^{2\phi(t,r)}dr^{2}-e^{2\mu(t,r)}d\theta^{2}
-e^{2\lambda(t,r)}dz^{2},
\end{equation}
where $\nu,~ \phi,~ \mu$ and $\lambda$ are arbitrary functions of
$t$ and $r$. The energy-momentum tensor for a perfect fluid is
given by
\begin{equation}
T_{ab}=[\rho(t,r)+p(t,r)]u_{a}u_{b}- p(t,r)g_{ab},\quad(a,b=0, 1,
2, 3),
\end{equation}
where $\rho$ is the mass density, $p$ is the pressure and $u_{a}$
is the four-velocity of the fluid element in the co-moving
coordinate system given as $u_{a}=(e^{\nu}, 0, 0, 0)$. The EFEs
for the line element (1) take the following form
\begin{eqnarray}
8\pi
G\rho&=&e^{-2\nu}(\mu_{t}\phi_{t}+\lambda_{t}\phi_{t}+\lambda_{t}\mu_{t})+
e^{-2\phi}(-\mu_{rr}+\mu_{r}\phi_{r}\nonumber\\
&-&\mu^{2}_{r}-\lambda_{rr}+\lambda_{r}\phi_{r}-
\lambda^{2}_{r}-\mu_{r}\lambda_{r}),\\
8\pi
Gp&=&e^{-2\nu}(-\mu_{tt}-\lambda_{tt}+\mu_{t}\nu_{t}+\lambda_{t}\nu_{t}-
\lambda_{t}\mu_{t}-\mu^{2}_{t}-\lambda^{2}_{t})\nonumber\\
&+&e^{-2\phi}(\mu_{r}\nu_{r}+\lambda_{r}\nu_{r}+\lambda_{r}\mu_{r}),\\
8\pi
Gp&=&e^{-2\nu}(-\phi_{tt}-\lambda_{tt}+\nu_{t}\phi_{t}+\lambda_{t}\nu_{t}-
\lambda_{t}\phi_{t}-\phi^{2}_{t}-\lambda^{2}_{t})\nonumber\\
&+&e^{-2\phi}(\nu_{rr}+\nu^{2}_{r}-\nu_{r}\phi_{r}+\lambda_{r}\nu_{r}
+\lambda_{rr}-\lambda_{r}\phi_{r}+\lambda^{2}_{r}),\\
8\pi
Gp&=&e^{-2\nu}(-\phi_{tt}-\mu_{tt}+\nu_{t}\phi_{t}+\mu_{t}\nu_{t}-
\mu_{t}\phi_{t}-\phi^{2}_{t}-\mu^{2}_{t})\nonumber\\
&+&e^{-2\phi}(\nu_{rr}+\nu^{2}_{r}-\nu_{r}\phi_{r}+\mu_{r}\nu_{r}
+\mu_{rr}-\mu_{r}\phi_{r}+\mu^{2}_{r}),\\
0&=&-\mu_{tr}-\lambda_{tr}+\mu_{t}\nu_{r}+\lambda_{t}\nu_{r}+\phi_{t}\mu_{r}+
\lambda_{r}\phi_{t}-\lambda_{r}\lambda_{t}-\mu_{t}\mu_{r}.
\end{eqnarray}
The conservation of energy-momentum tensor, ${T^{ab}}_{;b}=0$, gives
\begin{eqnarray}
\phi_{t}&=&-\frac{\rho_{t}}{\rho+p}-\mu_{t}-\lambda_{t},\\
\nu_{r}&=&-\frac{p_{r}}{\rho+p}.
\end{eqnarray}
For a cylindrically symmetric spacetime, the vector field $\xi$
can be written as
\begin{equation}
\xi^{a}\frac{\partial}{\partial
x^{a}}=h_{1}(t,r)\frac{\partial}{\partial
t}+h_{2}(t,r)\frac{\partial}{\partial r},
\end{equation}
where $h_{1}$ and $h_{2}$ are arbitrary functions. For $h_{2} =0$,
we obtain parallel case while for the orthogonal case $h_{1}=0$.
When both $h_{1}$ and $h_{2}$ are non-zero, we have tilted case
which is the most general. In this paper, we restrict ourselves to
investigate the parallel and orthogonal cases. The tilted case has
been explored separately [22].

A KSS vector $\xi$ satisfies the following conditions
\begin{eqnarray}
\pounds_{\xi} h_{ab}&=&2\delta h_{ab},\\
\pounds_{\xi} u_{a}&=&\alpha u_{a},
\end{eqnarray}
where $\alpha$ and $\delta$ are dimensionless constants and
$h_{ab}=g_{ab}-u_{a}u_{b}$ is the projection tensor. The similarity
transformation is characterized by the scale independent ratio,
$\alpha/\delta$, known as similarity index which gives rise to the
following possibilities according to $\delta\neq0$ or $\delta=0$.
\begin{itemize}

\item $\delta\neq0$.

For this possibility, the KSS vector takes the form
\begin{equation}
\xi^{a}\frac{\partial}{\partial x^a}=(\alpha
t+\beta)\frac{\partial}{\partial t}+r\frac{\partial}{\partial r}.
\end{equation}
If $\alpha=1$ ($\beta$ can be set to zero) then this case is
referred to as self-similarity of the first kind which corresponds
to a homothety.

When $\alpha=0$ ($\beta$ can be re-scaled to unity), this
corresponds to self-similarity of the zeroth kind.

When $\alpha\neq0,1$ ($\beta$ can be re-scaled to zero), this is
known as self-similarity of the second kind.

\item $\delta=0$.

In this case $\alpha\neq0$~ ($\alpha=1$ is possible, $\beta$ can be
re-scaled to zero), the self-similarity is referred to the infinite
kind.
\end{itemize}

For the parallel fluid flow, the KSS vector will take the form
\begin{equation}
\xi^{a}\frac{\partial}{\partial x^a}=f(t)\frac{\partial}{\partial
t},
\end{equation}
where $f(t)$ is an arbitrary function and self-similar variable is
$r$ for all kinds. The metric functions for the first, second,
zeroth and infinite kinds, respectively, will be
\begin{eqnarray}
&&\nu={\bar\nu(r)},\quad\phi=\ln t+{\bar\phi(r)},
\quad\mu=\ln t+{\bar\mu(r)},\quad\lambda=\ln t+{\bar\lambda(r)},\nonumber\\
&&\nu=(\alpha-1)\ln t+{\bar\nu(r)},\quad\phi=\ln
t+{\bar\phi(r)},\quad
\mu=\ln t+{\bar\mu(r)},\nonumber\\
&&\lambda=\ln t+{\bar\lambda(r)},\nonumber\\
&&\nu=-\ln t+{\bar\nu(r)},\quad\phi=\ln t+{\bar\phi(r)},\quad
\mu=\ln t+{\bar\mu(r)},\nonumber\\
&&\lambda=\ln t+{\bar\lambda(r)},\nonumber\\
&&\nu={\bar\nu(r)},\quad\phi={\bar\phi(r)},\quad\mu={\bar\mu(r)},\quad\lambda={\bar\lambda(r)}.
\end{eqnarray}
Hereafter we omit the bar for the sake of simplicity.
\begin{eqnarray}
&&\nu=\nu(r),\quad\phi=\ln t+\phi(r),
\quad\mu=\ln t+\mu(r),\quad\lambda=\ln t+\lambda(r),\nonumber\\
&&\nu=(\alpha-1)\ln t+\nu(r),\quad\phi=\ln t+\phi(r),\quad
\mu=\ln t+\mu(r),\nonumber\\
&&\lambda=\ln t+\lambda(r),\nonumber\\
&&\nu=-\ln t+\nu(r),\quad\phi=\ln t+\phi(r),\quad
\mu=\ln t+\mu(r),\nonumber\\
&&\lambda=\ln t+\lambda(r),\nonumber\\
&&\nu=\nu(r),\quad\phi=\phi(r),\quad\mu=\mu(r),\quad\lambda=\lambda(r).
\end{eqnarray}
If the KSS vector is orthogonal to the fluid flow, it becomes
\begin{equation}
\xi^a\frac{\partial}{\partial x^a}=g(r)\frac{\partial}{\partial
r},
\end{equation}
where $g(r)$ is an arbitrary function and self-similar variable is
$t$ for all  kinds. The corresponding metric functions for the
first, second, zeroth and infinite kinds, respectively, will take
the following form (after omitting the bar)
\begin{eqnarray}
&&\nu=\ln r+\nu(t),\quad\phi=\phi(t),
\quad\mu=\ln r+\mu(t),\quad\lambda=\ln r+\lambda(t),\nonumber\\
&&\nu=\alpha\ln r+\nu(t),\quad\phi=\phi(t),
\quad\mu=\ln r+\mu(t),\quad\lambda=\ln r+\lambda(t),\nonumber\\
&&\nu=\nu(t),\quad\phi=\phi(t),\quad
\mu=\ln r+\mu(t),\quad\lambda=\ln r+\lambda(t),\nonumber\\
&&\nu=\ln r+\nu(t),\quad\phi=-\ln
r+\phi(t),\quad\mu=\mu(t),\quad\lambda=\lambda(t).
\end{eqnarray}

Using coordinate transformation ${\bar r}={\bar r}(r)$ and ${\bar
t}={\bar t}(t)$, Eqs.(16) and (18) become
\begin{eqnarray}
&&\nu=\nu(r),\quad\phi=\ln t,
\quad\mu=\ln t+\mu(r),\quad\lambda=\ln t+\lambda(r),\nonumber\\
&&\nu=(\alpha-1)\ln t+\nu(r),\quad\phi=\ln t,\quad
\mu=\ln t+\mu(r),\nonumber\\
&&\lambda=\ln t+\lambda(r),\nonumber\\
&&\nu=-\ln t+\nu(r),\quad\phi=\ln t,\quad
\mu=\ln t+\mu(r),\nonumber\\
&&\lambda=\ln t+\lambda(r),\nonumber\\
&&\nu=\nu(r),\quad\phi=0,\quad\mu=\mu(r),\quad\lambda=\lambda(r),
\end{eqnarray}
and
\begin{eqnarray}
&&\nu=\ln r,\quad\phi=\phi(t),
\quad\mu=\ln r+\mu(t),\quad\lambda=\ln r+\lambda(t),\nonumber\\
&&\nu=\alpha\ln r,\quad\phi=\phi(t),
\quad\mu=\ln r+\mu(t),\quad\lambda=\ln r+\lambda(t),\nonumber\\
&&\nu=0,\quad\phi=\phi(t),\quad
\mu=\ln r+\mu(t),\quad\lambda=\ln r+\lambda(t),\nonumber\\
&&\nu=\ln r,\quad\phi=-\ln
r+\phi(t),\quad\mu=\mu(t),\quad\lambda=\lambda(t),
\end{eqnarray}
respectively.

\section{Parallel Perfect Fluid Case}

\subsection{Self-similarity of the First Kind}

For this kind, the EFEs imply that the quantities $\rho$ and $p$
can be written as
\begin{eqnarray}
8\pi G\rho&=&t^{-2}\rho(r),\\
8\pi Gp&=&t^{-2}p(r).
\end{eqnarray}
If the EFEs and the equations of motion for the matter field are
satisfied for $O[(t)^{-2}]$, we obtain a set of ODEs and thus
Eqs.(3)-(9) reduce to
\begin{eqnarray}
\nu^{'}&=&0,\\
\rho&=&3e^{-2\nu}+(-\mu^{''}-\mu^{'2}-\lambda^{''}
-\lambda^{'2}-\mu^{'}\lambda^{'}),\\
p&=&-e^{-2\nu}+(\lambda^{'}\mu^{'}),\\
p&=&-e^{-2\nu}+(\lambda^{''}+\lambda^{'2}),\\
p&=&-e^{-2\nu}+(\mu^{''}+\mu^{'2}),\\
0&=&\rho+3p,\\
0&=&-p^{'},
\end{eqnarray}
where prime represents derivative with respect to $r$. Equation (28)
is an equation of state (EOS) which gives along with Eq.(29)
$p=-\rho/3=-1$ and consequently, Eqs(24)-(27) lead to
$\lambda^{''}+\lambda^{'2}=0=\mu^{''}+\mu^{'2}$. Integration of
these equations yield the following solution
\begin{eqnarray}
&&\nu=c_{1},\quad\phi=0,\quad\mu=\ln(\xi-c_{3})+c_{2},
\quad\lambda=\ln(\xi-c_{5})+c_{4}\nonumber\\
&&p=-\rho/3=-1.
\end{eqnarray}
Equation (30) implies that $e^{\mu}=rc_{6}+c_{7}$ and
$e^{\lambda}=rc_{8}+c_{9}$. Using Eq.(30) with Eq.(25), we obtain
$c_{6}c_{8}=0$ which yields either $c_{6}=0$, or $c_{8}=0$, or
$c_{6}=0=c_{8}$.

In the first case, we set $c_{7}=1=c_{8}$. Also, $c_{9}=0$ by
re-defining the origin of $r$ , i.e., ${\bar r}=r+constant$. Thus
the resulting spacetime becomes
\begin{equation}
ds^2=dt^2-t^2(dr^2+d\theta^2+r^2dz^2).
\end{equation}
For the case, when $c_{8}=0$, we follow the same procedure as above
and obtain the same spacetime by interchanging $z$ and $\theta$. In
the last case, the resulting spacetime takes the form
\begin{equation}
ds^2=dt^2-t^2(dr^2+d\theta^2+dz^2).
\end{equation}
In all these cases, the spacetime corresponds to FRW metric.

\subsection{Self-similarity of the Second Kind}

Here the EFEs imply that the quantities $\rho$ and $p$ are of the
form
\begin{eqnarray}
8\pi G\rho&=&t^{-2}\rho_1(r)+t^{-2\alpha}\rho_{2}(r),\\
8\pi Gp&=&t^{-2}p_1(r)+t^{-2\alpha}p_{2}(r).
\end{eqnarray}
The corresponding set of ODEs become
\begin{eqnarray}
\nu^{'}&=&0,\\
\rho_{1}&=&-\mu^{''}-\mu^{'2}-\lambda^{''}-\lambda^{'2}-\mu^{'}\lambda^{'},\\
e^{2\nu}\rho_{2}&=&3,\\
p_{1}&=&\lambda^{'}\mu^{'},\\
e^{2\nu}p_{2}&=&(2\alpha-3),\\
p_{1}&=&\lambda^{''}+\lambda^{'2},\\
p_{1}&=&\mu^{''}+\mu^{'2},\\
0&=&\rho_{1}+3p_{1},\\
0&=&(2\alpha-3)\rho_{2}-3p_{2},\\
-p^{'}_{1}&=&0,\\
-p^{'}_{2}&=&0.
\end{eqnarray}
Using Eq.(35), we obtain from Eq.(37) $c^{2}_{1}=\rho_{2}/3$.
Equations (42) and (43) are two EOS where
$p_{1}=-\rho_{1}/3=constant$, and $p_{2}=(2\alpha-3)c^{2}_{1}$
respectively. Solving Eqs.(36), (38), (40) and (41), we obtain
\begin{eqnarray}
\mu^{''}+\mu^{'2}&=&p_{1},\\
\lambda^{''}+\lambda^{'2}&=&p_{1}.
\end{eqnarray}
Equations (38) and (46) lead to the following solution
\begin{eqnarray}
&&\nu=c_{1},\quad\phi=0,\quad\mu=c_{3}+\ln[\cosh[\sqrt{p_{1}}
(\xi+c_{2})]],\nonumber\\
&&\lambda=\ln[\sinh[\sqrt{p_{1}}(\xi+c_{2})]].
\end{eqnarray}
The resulting spacetime is
\begin{equation}
ds^2=dt^2-t^2(dr^2+\cosh^{2}[\sqrt{p_{1}}
(r+c_{2})]d\theta^2+\sinh^{2}[\sqrt{p_{1}}(r+c_{2})]dz^2).
\end{equation}
Equation (47) leads to the same spacetime by interchanging $\theta$
and $z$.

\subsection{Self-similarity of the Zeroth Kind}

For this kind, we set $\alpha=0$ in Eqs.(33)-(45). Equations (42)
and (43) give $p_{1}=-\rho_{1}/3=constant,~p_{2}=-\rho_{2}=-3$
which are equations of state. Solving Eqs.(35)-(45), we obtain the
same solution as given for the second kind.

\subsection{Self-similarity of the Infinite Kind}

For this kind, the metric functions of (1) are given by Eq.(19)
and a set of ODEs turn out to be
\begin{eqnarray}
\rho&=&-\mu^{''}-\mu^{'2}-\lambda^{''}-\lambda^{'2}-\mu^{'}\lambda^{'},\\
p&=&\mu^{'}\nu^{'}+\lambda^{'}\nu^{'}+\lambda^{'}\mu^{'},\\
p&=&\nu^{''}+\nu^{'2}+\lambda^{'}\nu^{'}
+\lambda^{''}+\lambda^{'2},\\
p&=&\nu^{''}+\nu^{'2}+\mu^{'}\nu^{'}
+\mu^{''}+\mu^{'2},\\
-p^{'}&=&\nu^{'}(\rho+p).
\end{eqnarray}
We take the following possibilities to solve the above set of
equations
\begin{eqnarray}
&&(1)\quad\nu'=\mu',\quad(2)\quad\nu'=\lambda',\nonumber\\
&&(3)\quad\lambda'=\mu',\quad(4)\quad\mu'=\nu'=\lambda'.
\end{eqnarray}
The possibilities (1) and (2) yield contradiction. The possibility
(3) gives the following solution
\begin{equation}
\phi=0,\quad\nu=c_{1},\quad\mu=\lambda=c_{2}\xi+c_{3},\quad
p=-\rho/3=constant,
\end{equation}
and the corresponding metric is
\begin{equation}
ds^{2}=dt^{2}-dr^{2}-e^{2c_{2}r}(d\theta^{2}+dz^{2}).
\end{equation}
The last possibility gives the following solution
\begin{equation}
\phi=0,\quad\nu=\mu=\lambda =c_{1}\xi+c_{2},\quad p=-\rho=constant,
\end{equation}
and the resulting spacetime is
\begin{equation}
ds^2= e^{2c_{1}r} (dt^2 -d\theta^2-dz^2)-dr^2.
\end{equation}

\section{Parallel Dust Case}

\subsection{Self-similarity of the First Kind}

For the dust case, we take $p=0$ in Eqs.(23)-(29). Solving
Eqs.(24)-(27), we obtain $\mu''+\mu'^2-\mu'\lambda'=0$ and
finally, we have the following solution
\begin{eqnarray}
&&\nu=c_{1},\quad\phi=0,\quad\mu=\ln(e^{2\xi}-c_{2})+c_{3}-\xi,\nonumber\\
&&\lambda=\ln(e^{2\xi}+c_{2})-\xi,\quad\rho=0=p.
\end{eqnarray}
The resulting spacetime is
\begin{equation}
ds^{2}= dt^2-t^{2}[dr^{2}+e^{-2r}[(e^{2r}-c_{2})^2
d\theta^2+(e^{2r}+c_{2})^2dz^2]].
\end{equation}

\subsection{Self-similarity of the Second Kind}

Setting $p_{1}=0=p_{2}$ in Eqs.(35)-(45), we obtain the same
solution as given by Eqs.(31) and (32) with
$\rho_{1}=0,~\rho_{2}=3c^2_{1}$ and $\alpha=3/2$, where $c_{1}$ is
the same as given in perfect fluid case.

\subsection{Self-similarity of the Zeroth Kind}

For $p_{1}=0=p_{2}$, we have contradiction from basic equations
for perfect fluid case.

\subsection{Self-similarity of the Infinite Kind}

When we replace $p=0$ in Eqs.(50)-(54), there arise three cases
from Eq.(54)
\begin{equation}
(a)\quad\nu^{'}=0,\quad\rho\neq0,\quad(b)\quad\rho=0,\quad\nu^{'}\neq0,\quad
(c)\quad\nu^{'}=0=\rho.
\end{equation}
The first case (a) yields contradiction and hence there is no
solution.

For the case (b), we take the four possibilities given in Eq.(55).
For the possibility (1), Eq.(51) further gives two more options
either $\mu'=0$ or $\mu'=-2\lambda'$. The first option gives
contradiction while the second option yields the following solution
\begin{eqnarray}
&&\nu=\mu=\frac{-2}{3}\ln(3\xi+c_{1}),\quad\lambda=c_{2}-\frac{1}{3}\ln(3\xi+c_{1}),
\quad\phi=0,\nonumber\\
&&\rho=p=0.
\end{eqnarray}
The corresponding metric becomes
\begin{equation}
ds^{2}=(3r+c_{1})^{4/3}(dt^{2}-d\theta^{2})-dr^{2}-(3r+c_{2})^{-2/3}dz^{2}.
\end{equation}
The possibility (2) yields the same solution by interchanging $z$
and $\theta$ and the possibility (3) leads to the following solution
\begin{eqnarray}
&&\lambda=\mu=c_{1},\quad\nu=c_{3}+\ln(\xi-c_{2}),\quad\phi=0,\quad\rho=p=0,\\
&&\lambda=\mu=\frac{-2}{3}\ln(3\xi+c_{1}),\quad\nu=c_{2}-\frac{1}{3}\ln(3\xi+c_{1}),
\quad\phi=0,\nonumber\\
&&\rho=p=0.
\end{eqnarray}
The corresponding metrics are
\begin{eqnarray}
ds^{2}&=&r^2dt^{2}-dr^{2}-d\theta^{2}-dz^{2},\\
ds^{2}&=&(3r+c_{2})^{-2/3}dt^{2}-dr^{2}-(3r+c_{1})^{4/3}(d\theta^{2}+dz^{2}).
\end{eqnarray}
For the last possibility (4), we have contradiction and hence there
is no solution.

The last case (c) gives the same solution as Eq.(30) with $\rho=0=p$
but the corresponding metrics are
\begin{eqnarray}
ds^{2}&=&dt^{2}-dr^{2}-d\theta^{2}-r^2dz^{2},\\
ds^{2}&=&dt^{2}-dr^{2}-r^2d\theta^{2}-dz^{2},
\end{eqnarray}
and \textit{Minkowski} spacetime.

\section{Orthogonal Perfect Fluid and Dust Cases}

Here the self-similar variable is $\xi=t$ in each kind. The EFEs
and the equations of motion for the perfect fluid of the first
kind gives the following set of ODEs
\begin{eqnarray}
\dot{\phi}&=&0,\\
(\rho+e^{-2\phi})&=&\dot{\mu}\dot{\lambda},\\
(p-3e^{-2\phi})&=&-\ddot{\mu}-\ddot{\lambda}
-\dot{\mu}\dot{\lambda}-\dot{\mu}^{2}-\dot{\lambda}^{2},\\
(p-e^{-2\phi})&=&-\ddot{\lambda}-\dot{\lambda}^{2},\\
(p-e^{-2\phi})&=&-\ddot{\mu}-\dot{\mu}^{2},\\
(\dot{\mu}+\dot{\lambda})(\rho+p)&=&-\dot{\rho},\\
\rho&=&p,
\end{eqnarray}
where dot represents derivative with respect to $t$. Equation (77)
gives an equation of state for this system of equations. Solving
Eqs.(71)-(77) by taking $\dot{\mu}=\dot{\lambda}$ which yields the
following solution
\begin{eqnarray}
&&\nu=0,\quad\phi=c_{1},\quad\mu=\lambda=-\frac{1}{4}\ln p+\ln
c_{2},\\
&&\ddot{p}=8p(3p+2),\quad p=p(t).
\end{eqnarray}
The resulting spacetime becomes
\begin{equation}
ds^{2}=dt^{2}-dr^{2}-\frac{r^2}{\sqrt{p}}(d\theta^{2}+dz^{2}).
\end{equation}
For the perfect fluid case of the second, zeroth and infinite kinds,
we obtain contradictions from the basic equations.

For the dust case (i.e., $p=0$) of the first kind, we obtain the
following solution
\begin{eqnarray}
&&\nu=0,\quad\phi=c_{1},\quad\mu=\ln(e^{2\xi}-c_{2})+c_{3}-\xi,\nonumber\\
&&\lambda=\ln(e^{2\xi}+c_{2})-\xi,\quad\rho=0=p,
\end{eqnarray}
and the corresponding metric is
\begin{equation}
ds^{2}=dt^{2}-dr^{2}-r^{2}e^{-2t}[(e^{2t}-c_{2})^2d\theta^{2}+(e^{2t}+c_{2})^2dz^{2}].
\end{equation}
In self-similarity of the second, zeroth and infinite kinds, we have
a contradiction and hence there is no solution for these kinds.

\section{Summary and Discussion}

We have evaluated the KSS perfect fluid and dust solutions of the
first, second, zeroth and infinite kinds when the KSS vector is
parallel and orthogonal to the fluid flow. In the parallel perfect
fluid case, the first kind gives only one vacuum solution while the
second kind gives one radiation solution. The zeroth kind yields the
same solution as for the second kind and the infinite kind yields
two independent vacuum solutions. In the orthogonal perfect fluid
case, there is only one stiff fluid solution for the first kind.

In the parallel dust case, the first kind yields only one vacuum
solution. The second kind yields the same solution as given in the
first kind for perfect fluid case while there exists no solution for
the zeroth kind. The infinite kind gives four independent vacuum
solutions out of which one is \textit{Minkowski} spacetime. For the
orthogonal dust case, the first kind provides only one vacuum
solution while in perfect fluid and dust cases there is no solution
for the second, zeroth and infinite kinds. Thus we obtain a total of
thirteen solutions out of which eleven are independent.

Here we give the correspondence of these solutions to the
well-known solutions available in the literature. The metrics
given by Eqs.(31) and (32) correspond to
\begin{equation}
ds^{2}=dt^{2}-a^{2}(t)[dr^{2}+f(r)d\theta^{2}+dz^{2}].
\end{equation}
This spacetime can be matched with FRW metric which has six KVs.
The spacetime (64) corresponds to a class of metrics [23]
\begin{equation}
ds^{2}=e^{\nu(r)}(dt^{2}-d\theta^{2})-dr^{2}-e^{\mu(r)}dz^{2}.
\end{equation}
The metrics given by Eq.(68) corresponds to a class of metrics
[23]
\begin{equation}
ds^{2}=e^{\nu(r)}dt^{2}-dr^{2}-e^{\mu(r)}(a^{2}d\theta^{2}+dz^{2}),
\end{equation}
where $\mu^{''}\neq0\neq\nu^{''}$. The spacetimes (84), (85) have
four KVs and belong to group $G_{4}=\langle X_{0}, X_{1}, X_{2},
X_{3}\rangle$.

The metric given by Eq.(57) belongs to a class of metrics [24]
\begin{equation}
ds^{2}=dt^{2}-dr^{2}-e^{r/b}(a^{2}d\theta^{2}+dz^{2}),
\end{equation}
where $b=1, 2,...,6$ and has 7 KVs while the metric (69)
corresponds to a class of metrics [24]
\begin{equation}
ds^{2}=dt^{2}-dr^{2}-a^{2}d\theta^{2}-e^{\lambda(r)}dz^{2},
\end{equation}
where $a=1, 2,...,6$. The spacetimes given by Eq.(70) correspond
to a class of metrics [24]
\begin{equation}
ds^{2}=dt^{2}-dr^{2}-a^{2}e^{\mu(r)}d\theta^{2}-dz^{2}.
\end{equation}
while Eqs.(67) correspond to a class of metrics [24]
\begin{equation}
ds^{2}=e^{\nu(r)}dt^{2}-dr^{2}-a^{2}d\theta^{2}-dz^{2}.
\end{equation}
The metrics given by Eqs.(87)-(89) have 6 KVs. The spacetime (59)
can correspond to the metric [24]
\begin{equation}
ds^{2}=(r/b)^2(dt^{2}-a^2d\theta^{2}-dz^{2})-dr^{2}
\end{equation}
which has 6 KVs. The metrics (49), (61), (80) and (82) do not
correspond to any solution in the literature.

It is interesting to mention here that the parallel case gives many
solutions while there was no solution for this case when dealing
with special metric [15]. The orthogonal case yields solutions only
in the first kind and contradictory results in all other kinds while
for the special metric [15], the infinite kind gives vacuum solution
and no solution in any other kind. The summary of the solutions are
presented below in tables 1-2.

\newpage

\textbf{Table 1.} Parallel Perfect Fluid KSS solutions.

\begin{center}
\begin{tabular}{|c|c|}
 \hline {\bf Self-similarity} & {\bf Solutions}
\\ \hline  First kind & solutions given by Eqs.(31),(32)
\\ \hline  Second kind & solution given by Eq.(49)
\\ \hline  Zeroth kind & solution given by Eq.(49)
\\ \hline  Infinite kind & solutions given by Eqs.(57),(59)
\\ \hline
\end{tabular}
\end{center}

\textbf{Table 2.} Parallel Dust KSS solutions.
\begin{center}
\begin{tabular}{|c|c|}
 \hline {\bf Self-similarity} & {\bf Solutions}
\\ \hline  First kind & solution given by Eq.(61)
\\ \hline  Second kind & solutions given by Eqs.(31),(32)
\\ \hline  Zeroth kind & None
\\ \hline  Infinite kind  & solutions given by
Eqs.(64),(68),(69),(70)
\\ \hline
\end{tabular}
\end{center}
There is only one solution each for the orthogonal perfect fluid and
dust cases of the first kind given by Eqs.(80) and (82)
respectively.

\end{document}